# Slime Mould Inspired Generalised Voronoi Diagrams with Repulsive Fields

Jeff Jones

*Centre for Unconventional Computing, University of the West of England,*
*Coldharbour Lane, Bristol, BS16 1QY, UK*
*jeff.jones@uwe.ac.uk*

Andrew Adamatzky

*Centre for Unconventional Computing, University of the West of England,*
*Coldharbour Lane, Bristol, BS16 1QY, UK*
*andrew.adamatzky@uwe.ac.uk*

The Voronoi diagram is a widely used partition of space which may be implemented in a variety of unconventional computing prototypes, sharing in common the uniform propagation of information via fronts emanating from data point sources. The giant single-celled amoeboid organism *Physarum polycephalum* constructs minimising transport networks but can also approximate the Voronoi diagram using two different mechanisms. In the first method Voronoi bisectors are represented by deformation of a pre-existing plasmodial network by repellent sources acting as generating points. In the second method generating points act as inoculation sites for growing plasmodia and Voronoi bisectors are represented by vacant regions before the plasmodia fuse. To explore the behaviour of minimising networks in the presence of repulsion fields we utilise a computational model of *Physarum* as a distributed virtual computing material. We characterise the different types of computational behaviours elicited by attraction and repulsion stimuli and demonstrate the approximation Voronoi diagrams using growth towards attractants, avoidance of repellents, and combinations of both. Approximation of Voronoi diagrams for point data sources, complex planar shapes and circle sets is demonstrated. By altering repellent concentration we found that partition of data sources was maintained but the internal network connectivity was minimised by the contractile force of the transport network. To conclude, we find that the repertoire of unconventional computation methods is enhanced by the addition of stimuli presented by repellent fields, suggesting novel approaches to plane-division, packing, and minimisation problems.

## 1. Introduction

The Voronoi diagram of a set of $n$ points in the plane is the subdivision of the plane into $n$ cells so that every location within each cell is closest to the generating point within that cell. Conversely the bisectors forming the diagram are equidistant from the points between them. Voronoi diagrams are useful constructs historically applied in diverse fields as computational geometry, biology, epidemiology, telecommunication networks and materials science. Efficient computation of the Voronoi diagram may be achieved with a number of classical algorithms [Fortune, 1987; De Berg *et al.*, 2008] and are also amongst prototypical ap-plications solved by chemical reaction-diffusion non-classical computing devices [Tolmachiev & Adamatzky, 1996; de Lacy Costello *et al.*, 2004b]. Non-classical approaches are typically based upon the intuitive notion of uniform propagation speed within a medium, emanating from the source nodes. The bisectors of the diagram are formed where the propagating fronts meet, visualised, for example in chemical processors, by the lack of precipitation where the fronts merge [de Lacy Costello *et al.*, 2004b]. Voronoi diagrams can be generated in other physical systems by generalising the propagation mechanism and visualisation of the bisectors and have been implemented in a number of different media including reaction-diffusion chemical processors [Tolmachiev & Adamatzky, 1996; de Lacy Costello *et al.*, 2004b], planar silicon [Asai *et al.*, 2005], crystalline phase change materials [Adamatzky, 2009a], and gas discharge systems [Zanin *et al.*, 2002]. In living systems approximation of Voronoi diagrams may be achieved by inoculating a chosen organism or cell type at the source points on a suitable substrate. Outward growth from the inoculation site corresponds to the propagative mechanism and regions where the colonies or cells meet correspond to bisectors of the diagram.

The true slime mould *Physarum polycephalum* has also been shown to approximate the Voronoi diagram by two different methods, based on its interactions with environments containing repellents [Shirakawa & Gunji, 2010] and attractants [Adamatzky, 2010b]. *Physarum* is a single-celled organism visible to the naked eye. In the plasmodium stage of its complex life-cycle it forms a multi-nucleate syncytium bound by a single membrane, growing up to the square metre scale in size. The growing plasmodium extends outwards as it forages for nutrients, streaming forwards with a characteristic pulsatile progression at up to 1cm/h [Wohlfarth-Bottermann, 1986]. The plasmodium adapts its body plan to form a protoplasmic tube network which transports microscopic nutrient fragments between different parts of the plasmodium. Under controlled conditions the growth patterns of the plasmodium are strongly affected by nutrient concentration [Takamatsu *et al.*, 2009]. At low concentrations (for example damp filter paper or non-nutrient agar) the growth is by extension of pseudopodium-like processes which project towards the nutrients (Fig. 1a). When *Physarum* is inoculated on a nutrient rich substrate, such as oatmeal agar, florid radial growth is observed and the plasmodium propagates outwards with a circular growth front (Fig. 1b-d).

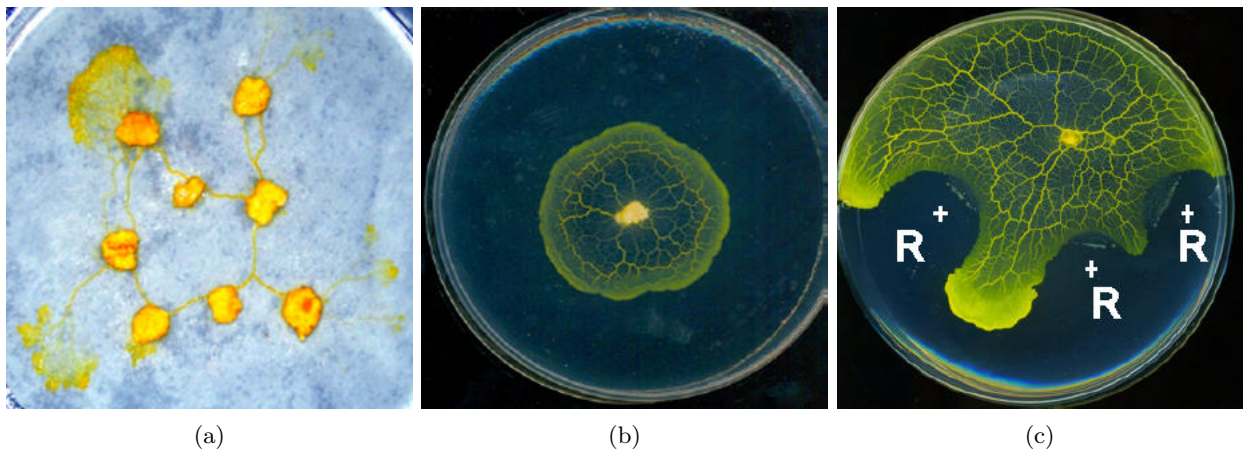

(a)            (b)            (c)

Fig. 1. Propagation of *Physarum* under the influence of attractants and repellents. (a) dendritic growth of plasmodium transport network (yellow) on damp filter paper with oat flake stimuli approximates Spanning Tree, (b) radial growth from a single site of inoculation on a nutrient rich substrate, (c) avoidance of repellents (salt grains located at crosses) which have diffused into neighbouring regions 'R'.

The protoplasmic tube network is comprised of an external membrane, the plasmalemma, surrounding concentric layers of outer gel-like ectoplasm and an inner watery, or sol-like, inner endoplasm core [Achenbach *et al.*, 1979]. The ectoplasm contains fibrils composed of actin and myosin complexes which are arranged primarily longitudinally and circumferentially, adhering to invaginations in the plasmalemma. Internal metabolic processes and external stimulation (for example, chemoattractants, chemorepellents, temperature) [Ueda *et al.*, 1975], [Matsumoto *et al.*, 1986] cause fluctuations in the concentration of in-tracellular compounds (including ATP, ADP, Ca2$^+$, cAMP, cGMP, NADH, H$^+$) [Hejnowicz & Wohlfarth-Bottermann, 1980; Ridgway & Durham, 1976; Ueda *et al.*, 1986; Yoshimoto *et al.*, 1981]. The actomyosin complexes within the ectoplasm contract and relax in response to changing concentrations of intracellular compounds, shortening the length of the fibrils and compressing the inner endoplasm core [Hatano & Oosawa, 1966; Isenberg & Wohlfarth-Bottermann, 1976]. Protoplasmic veins contract both radially and longitudinally [Ishigami, 1986; Wohlfarth-Bottermann, 1979] and the hydrostatic pressure induced by the contractions passively transports the endoplasmic sol along the protoplasmic tube network [Kamiya, 1950]. The oscillatory activity results in a to-and-fro transport of protoplasm known as shuttle streaming at a velocity of up to 1.3 mm/s [Kamiya & Kuroda, 1958] and protoplasm is transported to the leading edge of the growth front.

The computational behaviour of *Physarum* was stimulated by the findings of Nakagaki et al. who reported the ability of the *P. polycephalum* plasmodium to solve a simple maze problem. When multiple nutrient sources were placed within a large plasmodium the organism adapted its morphology to form efficient paths (in terms of a trade-off between overall distance and resilience to random damage) between the food sources [Nakagaki *et al.*, 2004],[Nakagaki & Guy, 2007],[Nakagaki *et al.*, 2007]. Subsequent research has demonstrated that the plasmodium successfully approximates spatial representations of various graph problems [Nakagaki *et al.*, 2004; Shirakawa *et al.*, 2009; Adamatzky, 2008; Jones, 2010a], combinatorial optimisation problems [Aono & Hara, 2007, 2008; Ozasa *et al.*, 2010; Jones, 2011], construction of logic gates and adding circuits [Tsuda *et al.*, 2004; Jones & Adamatzky, 2010; Adamatzky, 2010e], and spatially represented logical machines [Adamatzky, 2007; Adamatzky & Jones, 2010]. *P. polycephalum* has also been used directly as a means to achieve distributed robotic control [Tsuda *et al.*, 2007], direct robotic transport and guidance [Adamatzky, 2010a, 2009b], robotic manipulation [Adamatzky & Jones, 2008; Adamatzky, 2010c], and also indirectly as an inspiration for robotic amoeboid movement [Umedachi & Ishiguro, 2007; Umedachi *et al.*, 2010; Jones & Adamatzky, 2012].

*Physarum* has been used to approximate the Voronoi diagram using two different methods based on interactions within environments containing repellents and attractants and may be considered as a living form of material computation. In this paper we examine the use of attraction and repulsion fields in spatially represented unconventional computation. In section 2 we describe in more detail the different methods of utilising *Physarum* for constructing Voronoi diagrams. A computational perspective on the role of attractants and repellents is given in section 3 which is used to inform modelling approaches. A spatially represented computational model of *Physarum* is described in section 4. In section 5 we examine how different inoculation arrangements and the response to repulsion fields affect the patterning of the model. We explore the effects of repellent size and concentration on Voronoi diagram construction and show hybrid diagram constructs which take advantage of the model's innate minimising behaviour with a simultaneous response to repulsion fields. In section 6 we summarise the repertoire of spatial combination using the field approach and suggest some computational tasks which may be amenable to this approach.

## 2. Approximation of Voronoi diagram by *Physarum polycephalum*

### 2.1. *Physarum Represents Voronoi Bisectors*

The method of using *Physarum* to approximate Voronoi diagrams by avoidance of chemorepellents was described in [Shirakawa *et al.*, 2009; Shirakawa & Gunji, 2010]. In this method a fully grown large plasmodium was first formed in a circular arena. Then repellent sources were introduced onto the plasmodium. The circular border of the arena was surrounded by attractants to maintain connectivity of the plasmodium network. The plasmodium then adapted its transport network to avoid the repellents whilst remain connected to the outer attractants, approximating the Voronoi diagram (Fig. 2).

### 2.2. *Absence of Physarum Represents Bisectors*

Computation of Voronoi diagram may also be achieved by non-repellent methods. This method is proposed in [Adamatzky, 2010b] where plasmodia of *Physarum* are inoculated at node sites on a nutrient-rich agar substrate (Fig. 3a). Attracted by the surrounding stimuli the plasmodia grow outwards in a radial pattern but when two or more plasmodia meet they do not immediately fuse (Fig. 3b). There is a period where the growth is inhibited (presumably via some component of the plasmodium membrane or slime capsule) and the substrate at these positions is not occupied, approximating the Voronoi diagram (diagram computed by classical method is shown in Fig. 3c). The position of the growth fronts remains stable before complete fusion eventually occurs (Fig. 3d). Also of note is the growth emanating from the lower left node in which growth of the plasmodium inoculated at this point is slower (Fig. 3c, circled region). The relatively slow progression of the growth front from this flake allows growth fronts from nearby flakes to encroach into the area and thus distort the position where the fronts meet, compared to the bisectors in the classical method. This is similar to the situation noted in chemical systems [de Lacy Costello *et al.*, 2004a] when delayed instantiation of chemical diffusion fronts results in less freedom of propagation, approximating the additively weighted Voronoi diagram.

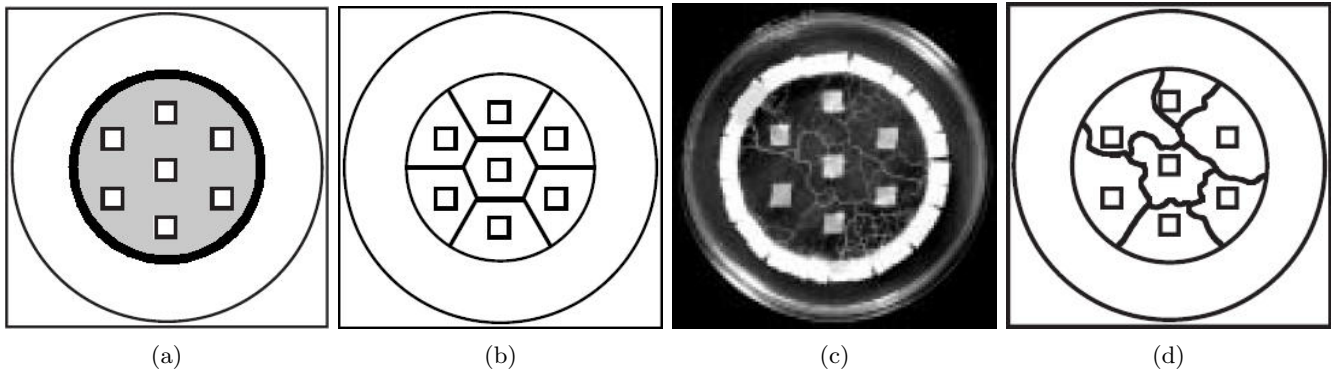

Fig. 2. Approximation of Voronoi diagram by *Physarum* using the repellent method: (a) Configuration of arena showing nutrient attractants (thick ring), Repellents (squares) and inoculation area of plasmodium (grey), (b) Voronoi diagram for this configuration, (c) example approximation of Voronoi diagram formed by redistribution of *Physarum* transport network, (d) Enhanced visualisation of transport network approximation (Images courtesy of Tomohiro Shirakawa).

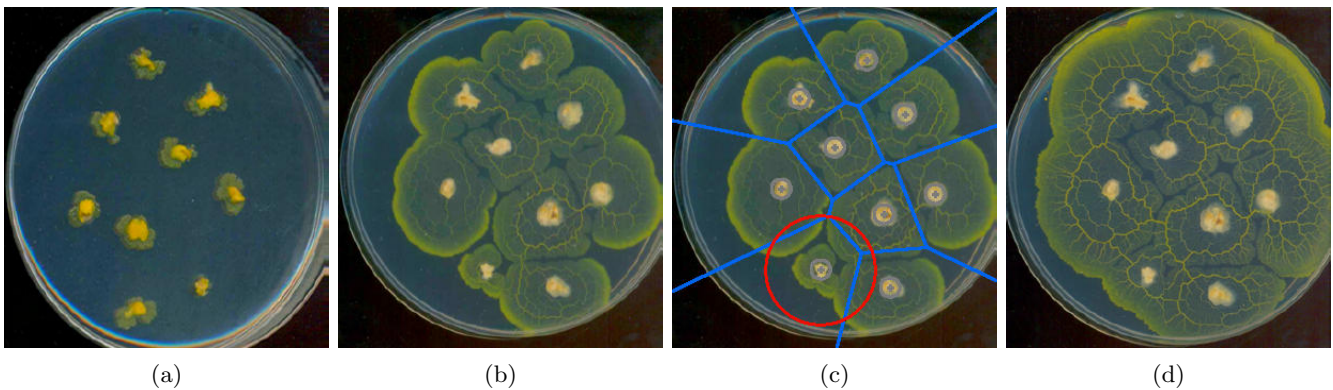

Fig. 3. Approximation of Voronoi diagram by *Physarum* using the merging fronts method: (a) Inoculation sites represent data planar points to be subdivided by edges of Voronoi diagram, (b) propagation of growing plasmodia at 10h. Bisectors of Voronoi diagram are represented by loci of substrate not occupied by plasmodium, (c) Plasmodium with overlaid Voronoi Diagram (blue) computed by classical method, (d) Merging of growth fronts at 16h. Images from [Adamatzky, 2010b].

## 3. Computational Perspectives on Slime Mould Computation

*Physarum* may be considered as a living form of material computation, i.e. a material whose deformation and adaptation in response to environmental stimuli approximates a range of spatially represented computational geometry problems. Its behaviour results in maximising the area explored, using minimal connectivity (the exploration vs exploitation trade-off [Gunji *et al.*, 2011]). *Physarum* does not provide exact solutions to these problems and its behaviour is somewhat unpredictable. It does not exhibit the predictable material evolution seen in, for example, soap films, since the organism is concerned merely with survival, rather than solving externally applied problems.

By interpreting the innate spatial behaviour of slime mould and its environmental interactions from a computational perspective it may be possible to gain some clues as to the development of future morphological computation substrates and means of externally influencing their behaviour. Note that we are only concerned with classifying the direct spatially represented computational behaviour of *Physarum*. Other computational interpretations may be possible based upon encoding of information via its oscillatory phenomena [Adamatzky. & Jones, 2011], or network patterning as a process calculus [Schumann & Adamatzky, 2011] or graph-based computing machine [Adamatzky, 2007].

The plasmodium migrates due to the hydrostatic pressure exerted on the plasmodium by oscillatory contractions within the material, causing protrusion from softer parts of the membrane. In the absence of nutrient stimuli the migration is dendritic with pseudopodium-like processes extending forwards at the growth front [Takamatsu *et al.*, 2009]. Flux canalisation within the *Physarum* tube network has been suggested as a mechanism of network adaptation and minimisation [Tero *et al.*, 2008; Nakagaki *et al.*, 2004] whereby shorter tubes with high flux are rewarded (by becoming thicker and increasing their conductance) and longer tubes with less flux are penalised (by becoming thinner and decreasing their conductance). This mechanism can account for network adaptation when nutrient sources are added to an arena containing a pre-existing plasmodium. Thus the 'default' behaviour of the *Physarum* plasmodium are exploration and minimisation processes (Fig. 4). In the absence of any stimuli and nutrient energy these process continues until, starved of nutrients, the plasmodium reverts to the dormant sclerotium stage of its lifecycle. Regeneration to a plasmodium may occur if moisture and nutrients are subsequently added.

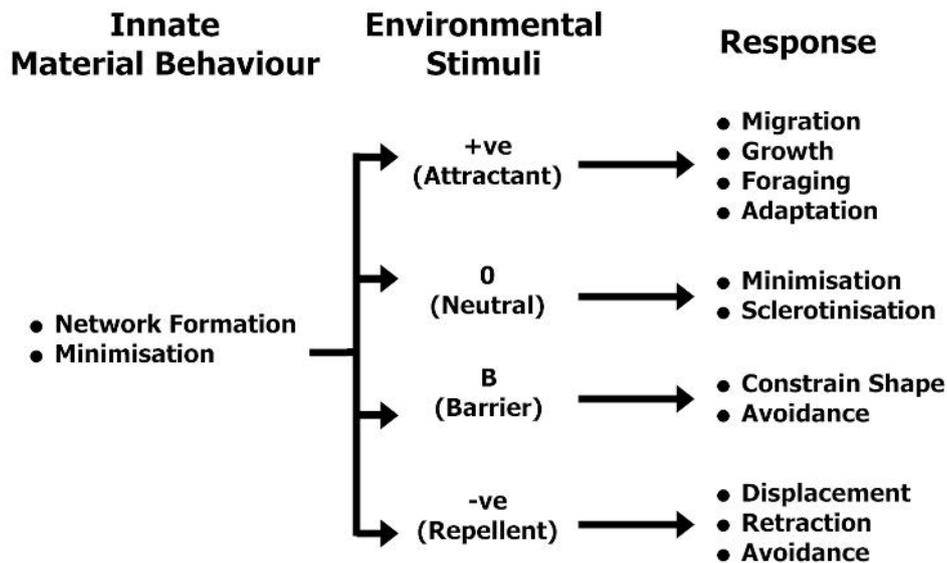

Fig. 4. Computational perspective of innate *Physarum* behaviour and response to attractant, neutral and repellent stimuli

External environmental stimuli may be attractive (+ve), neutral (0ve), or repulsive (-ve) to the plasmodial network. If we assume neutral stimuli have no effect on the network (other than the gradual reduction in size due to lack of nutrients) we may consider attractant and repellent stimuli in isolation.

Chemoattractant gradients diffusing from +ve stimuli invoke growth of the plasmodium towards the stimuli. Depending on the strength of the gradient the extension is via pseudopodium-like processes (weak stimuli, Fig. 1a) or via a fan-like growth front (stronger stimuli). A special case exists if the plasmodium is surrounded by very high concentration stimuli. In this case the growth will be outward in all directions, generating a radial pattern (Fig. 1b). It is important to note that the radial outward growth of the plasmodium in this instance is not due to repulsion at the inoculation site. The plasmodium is instead *pulled* outwards by the surrounding stimuli.

In terms of -ve stimuli it may be difficult to state if the cause of the response is due to preferential avoidance of the stimulus, for example by migrating away from it, or whether the stimulus has a less subtle effect, such as destroying the plasmodium in regions where it touches the stimulus. Nevertheless, it can be said that the general response to -ve stimuli is to move away from the stimulus. Special cases exist where the environment presents unfavourable conditions, rather than actual an actual repellent stimulus. For example a migrating plasmodium will preferentially avoid dry regions placed on agar substrate (e.g. acetate film). In this case the stimulus acts as an barrier or obstacle, even though no -ve stimulus diffuses from the source, and can be used to constrain movement of the organism [Nakagaki *et al.*, 2000] or guide its migration.

The interaction between *Physarum* and its environment is a complex modification of the innate patterning of the organism. We may interpret *Physarum* as a deformable material which can be deformed towards +ve attractant stimuli and away from -ve repellent stimuli. As it is difficult to control *Physarum* precisely using these methods we must turn to modelling to investigate the effects of combining both +ve and -ve stimuli.

## 4. A Virtual Material Modelling Approach

We use the multi-agent approach introduced in [Jones, 2010b] which was used to create emergent and minimising transport networks. This model uses simple, particle-like agents residing on a diffusive lattice. The particles collectively behave as a virtual material which has similar pattern formation and evolution characteristics in common with the plasmodium. Although the model is very simple in its assumptions and implementation it is capable of reproducing some of the phenomena of spontaneous network formation, foraging behaviour, oscillatory behaviour, bi-directional shuttle streaming, and network adaptation seen in *Physarum* using only simple, local microscopic functionality to generate collective macroscopic emergent behaviour.

Each particle represents a hypothetical unit of *Physarum* plasmodium in its contractile gel phase or its flowing sol phase. Collectively the population can approximate the effect of chemoattractant gradients on the plasmodium membrane (sensory behaviour) and the flow of protoplasmic sol within the plasmodium (motor behaviour). The collective particle *positions* at any instance in time correspond to a static snapshot of transport network structure, whilst collective particle *movement* corresponds to protoplasmic flow within the network.

The multi-agent particle approach to modelling the behaviour of *Physarum* plasmodium uses a population of mobile particles with very simple behaviours, residing within a 2D diffusive lattice. The lattice (where the features of the environment are mapped to grey-scale values in a 2D image) stores particle positions and the concentration of a local factor which we refer to generically as chemoattractant. The 'chemoattractant' factor actually represents the hypothetical flux of sol within the plasmodium. Free particle movement represents the sol phase of the plasmodium. Particle positions represent the fixed gel structure (i.e. global pattern) of the plasmodium. The particles act independently and iteration of the particle population is performed randomly to avoid any artifacts from sequential ordering. The behaviour of the particles occurs in two distinct stages, the sensory stage and the motor stage. In the sensory stage, the particles sample their local environment using three forward biased sensors whose angle from the forwards position (the sensor angle parameter, SA), and distance (sensor offset, SO) may be parametrically adjusted (Fig. 5a). The offset sensors represent the overlapping and intertwining filaments and generate local coupling of sensory inputs and movement to represent the protoplasmic tubes forming the transport networks and the mass of the plasmodium itself. The overlapping sensors loosely correspond to the coupling caused by the cross-linking of actin filaments in motile cells [Pollard & Borisy, 2003]. The SO distance is measured in pixels and a minimum distance of 3 pixels is required for strong local coupling to occur. During the sensory stage each particle changes its orientation to rotate (via the parameter rotation angle, RA) towards the strongest local source of chemoattractant (Fig. 5b). After the sensory stage, each particle executes the motor stage and attempts to move forwards in its current orientation (an angle from 0–360 degrees) by a single pixel forwards. Each lattice site may only store a single particle and particles deposit chemoattractant into the lattice only in the event of a successful forwards movement. If the next chosen site is already occupied by another particle the default (i.e. non-oscillatory) behaviour is to abandon the move and select a new random direction.

We use an extension of the basic particle model to include plasmodium growth and adaptation (growth and shrinkage of the collective) which is implemented as follows. If there are 1 to 10 particles in a $9 \times 9$ neighbourhood of a particle, and the particle has moved forwards successfully, the particle attempts to divide into two if there is an empty location in the immediate $3 \times 3$ neighbourhood surrounding the particle. If there are 0 to 24 particles in a $5 \times 5$ neighbourhood of a particle the particle survives, otherwise it is annihilated. The frequency at which the growth/shrinkage of the population is executed determines a turnover rate for the particles. For these experiments we used a high turnover rate (every three scheduler steps) to ensure a resilient network structure in response to changing presence and concentrations of +ve and -ve stimuli.

Particles were inoculated at the border of planar shapes in the arena (white regions). Particle sensor offset (SO) was 5 pixels except where explicitly stated. Angle of rotation (RA) and sensor angle (SA) were both set to 60 degrees in all experiments. Agent forward displacement was 1 pixel per step and particles moving forwards successfully deposited 1 units into the diffusive lattice, resulting in a temporary storage of agent movement history in the diffusion map. The emergent transport networks are represented in the results by the configuration of the particle collective, shown as a spatial map of particle positions.

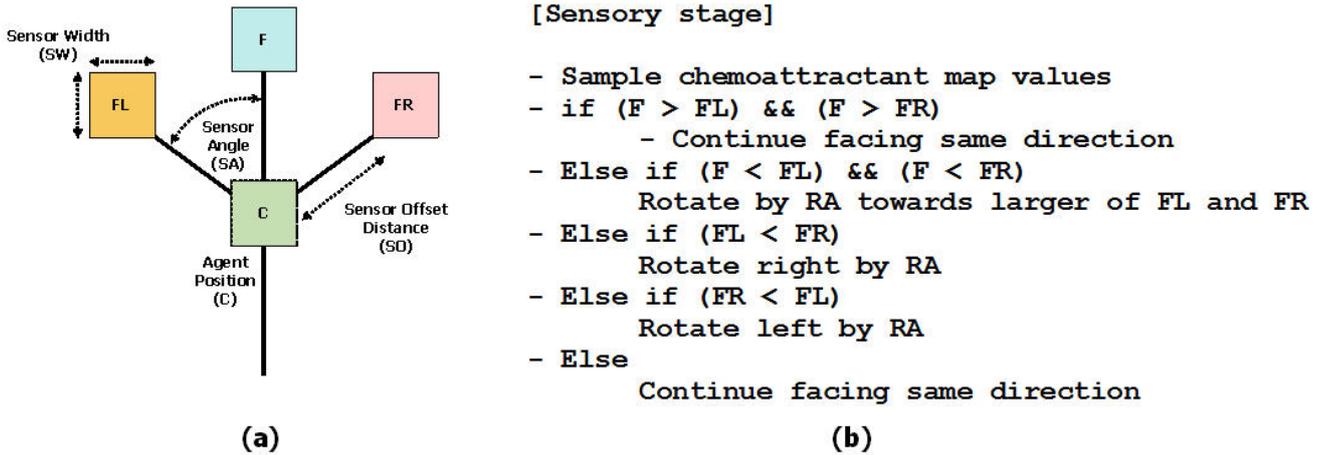

Fig. 5. Particle morphology and simplified algorithm. (a) Morphology showing agent position 'C' and sensor positions (FL, F, FR), (b) Algorithm for particle sensory stage.

### 4.1. *Representation of Attractant and Repulsion Diffusion Field*

The spatial configuration the environment is represented by a greyscale coded image isomorphic to the 2D lattice (Fig. 6a). Particular greyscale values represent uninhabitable boundaries, vacant areas (within which the population can grow, move and adapt the collective morphology) and locations of nutrients. Both attractant and repellent sources are represented by sources of chemoattractant concentration in the same diffusive lattice. Attractant sources are denoted by those which have a positive ($> 0$) value and repellent sources are represented by values $< 0$. The resulting profile of chemoattractant gradients may be interpreted as a 3D stimulus landscape (Fig. 6b)

Environmental stimuli are diffused by means of a simple mean filter kernel of size $D_w$, (typically 5×5). The diffusion operator is applied to all cells simultaneously via pseudo-parallelism methods. The diffusing chemoattractant values may be damped by the value $D_d$ (typically 0.1) to adjust the concentration of the diffusion gradient away from the nutrient source. (mean of $D_w$, multiplied by $(1 - D_d)$, smaller values of $D_d$ resulting in less damping of diffusion). Differences in stimulus concentration (greyscale value in the configuration map) and stimulus area (the size of nutrient source), affect both the steepness, and propagation distance of the diffusion gradient.

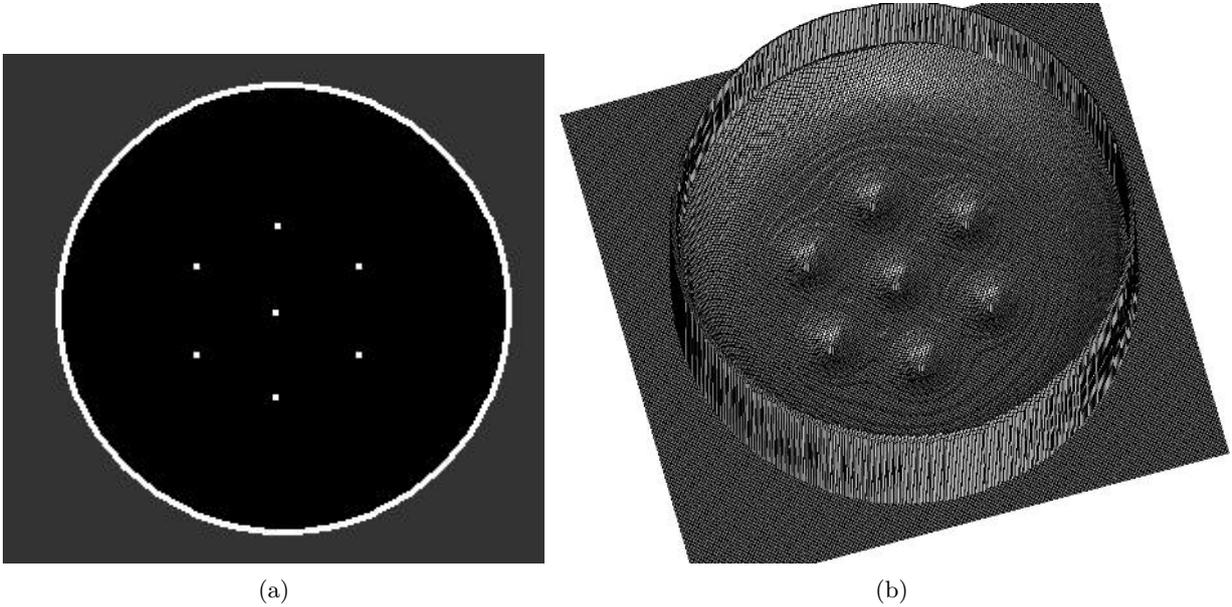

Fig. 6. Representation of model arena configuration as attractant and repellent gradient field. (a) Configuration of model arena showing circular attractant boundary and repellent sources, (b) 3D visualisation of concentration gradient field.

## 5. Results

### 5.1. *Interactions Between Attractant and Repellent Fields*

Fig. 7 shows the effect of -ve stimuli on path direction. A transport network attached by two +ve stimuli at each end (Fig. 7a, attractants circled) is deflected by the nearby placement of -ve stimuli (Fig. 7b, c). When the -ve stimuli are removed the path returns to its original configuration. Path deflection may also be cumulative if successive -ve stimuli are added to the arena (Fig. 8).

As with attractants, the concentration of -ve stimuli affects their influence on nearby particle networks as the stimuli diffuse into the gradient field. Fig. 9 illustrates the effect of concentration level as a circular minimising path (formed by self-assembly of particles introduced at random in the arena) is constrained by the repellent stimuli emanating from the central node. As the concentration increases, the -ve stimuli diffuse further from the source and the circular path is deformed outwards (Fig. 9a-d). If the concentration is subsequently reduced, the network path again contracts, limited by the diffusion distance from the source. Fig. 9e plots the increasing distance (circle radius) from the diffusion source as the -ve stimulus concentration increases. Separate plots for increasing values of particle sensor offset (SO) distances are given.

### 5.2. *Approximation of Voronoi diagram using repulsion method*

A population of particles was introduced into a circular arena. The interior of the arena was patterned with seven repellent sources and the border of the arena was patterned with attractant sources, reproducing the experimental pattern used in [Shirakawa & Gunji, 2010]. The uniform distribution of particles was repelled by the field emanating from the repellents and attracted by the sources at the border (Fig. 10), approximating the Voronoi diagram of the repellent sources.

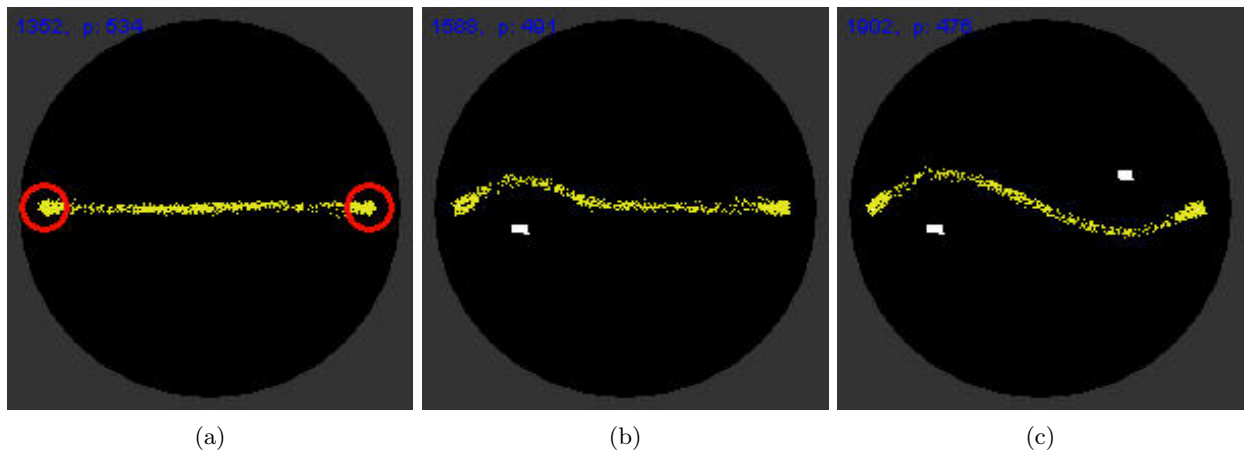

Fig. 7. Transport network path deflection by placement of repellent sources. (a) Transport network (yellow) in circular arena between two attractants (circled), (b) repellent source introduced below path distorts path away from repellent, (c) diffusion from second repellent source above the path deflects the path downwards.

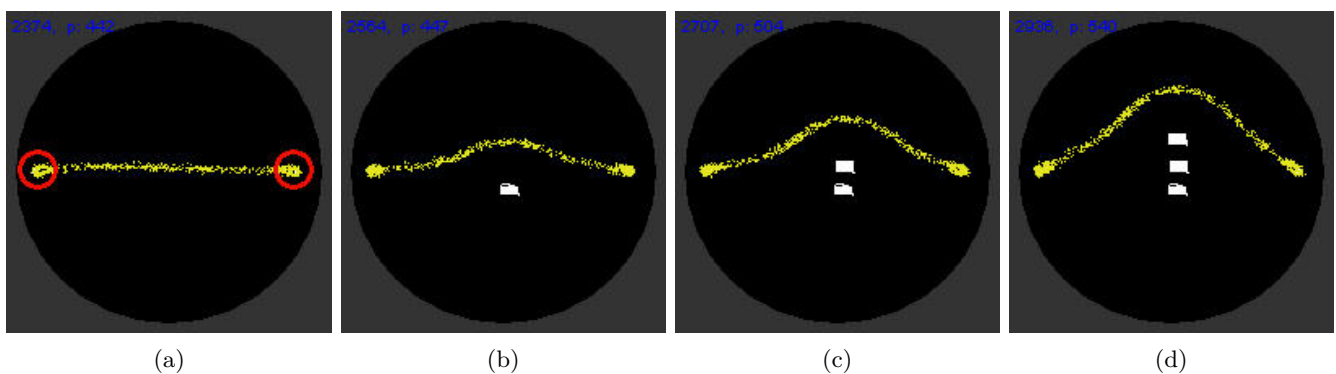

Fig. 8. Summation of transport network path deflection by placement of repellent sources. (a) Transport network (yellow) in circular arena between two attractants (circled), (b-d) addition of repellent sources deflects path upwards, causing steeper deflection as more repellents are added.

### 5.3. *Approximation of Voronoi diagram using merging method*

To approximate the Voronoi by the merging method in [Adamatzky, 2010b] a small population of particles were initialised at locations on a simulated nutrient rich background substrate corresponding to Voronoi source nodes. The strong stimulation from the high-concentration background generated radial growth and the Voronoi Diagram is approximated in the model at regions where the separate growth fronts fuse. Note that the model does not explicitly incorporate the inhibition at the touching growth fronts. In this case the Voronoi bisector position is instead indicated by the model by the increase in network density at the bisector position (Fig. 11b). This approximation is not satisfactory from a computational perspective since it requires subjective visual evaluation and interpretation.

### 5.4. *Approximation of Voronoi diagram using combined repulsion and merging method*

Can we combine the repulsion and merging methods to compute the Voronoi diagram? To test this question we use a more challenging problem where the sources for the bisectors are planar shapes or curves instead of single point sources, requiring the use of algebraic curves in classical algorithms. Chemical reaction diffusion processors are, however, capable of approximating the Voronoi diagram of complex shapes [Adamatzky & de Lacy Costello, 2003]. To see if the model plasmodium can approximate planar shape Voronoi diagrams and to see if the repulsion and merging methods can be used simultaneously, a population of model particles was initialised at the borders of repellent planar shapes in a simulated arena. The circular arena border was also configured as a repellent source and the remainder of the arena was configured to simulate a nutrient-rich background substrate (light grey regions) which was consumed on contact with the model population, Fig.12a.

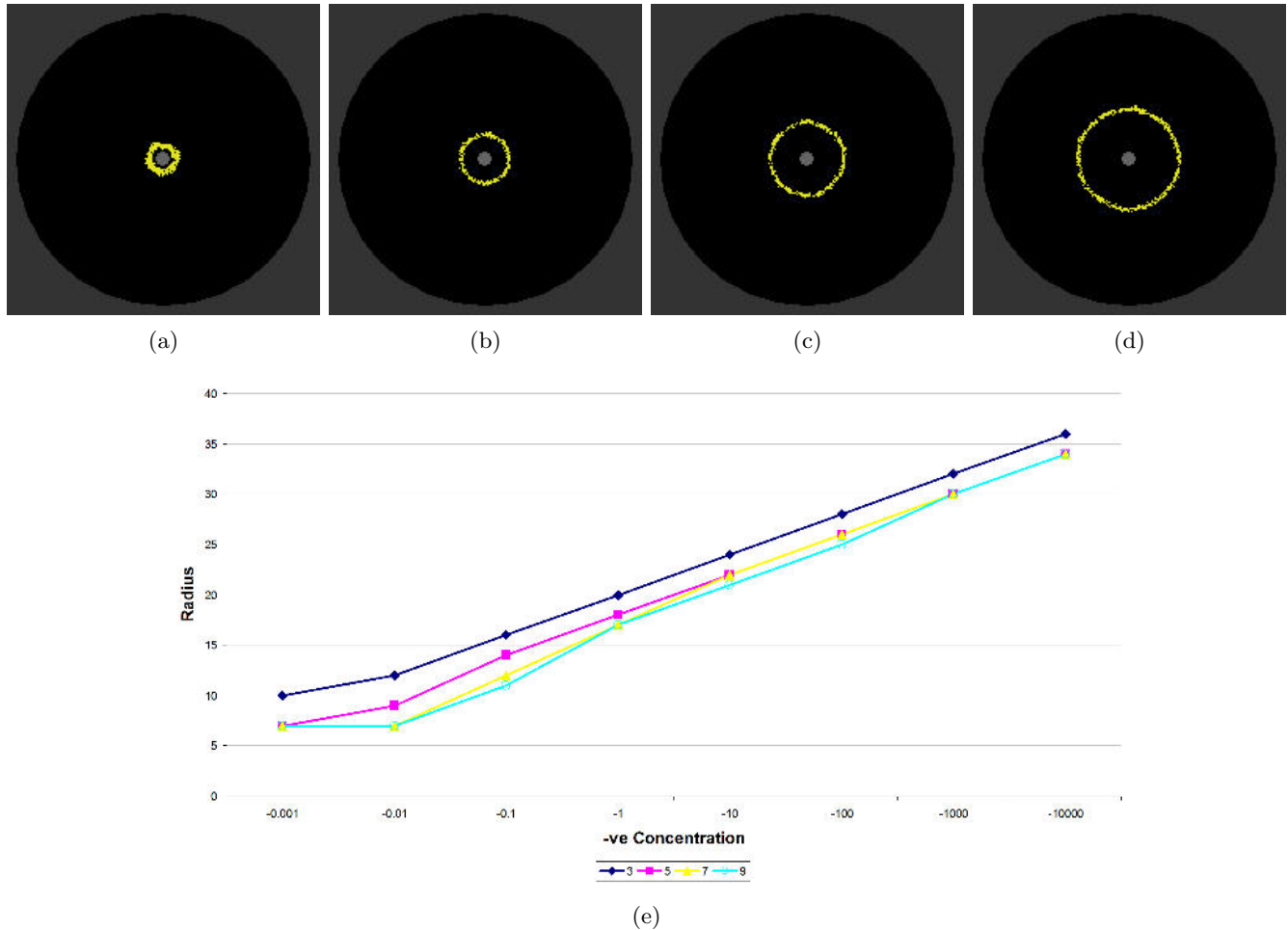

Fig. 9.   Effect of repellent concentration on contraction of transport network. (a) Model initialised in circular arena with central repellent and network (yellow) contracts around central point, (b-d) as repellent concentration increases the collective is repelled by the diffusion front from the circle, (e) plot of increasing radius from centre as repellent concentration increases. Four experiments shown with different SO distance.

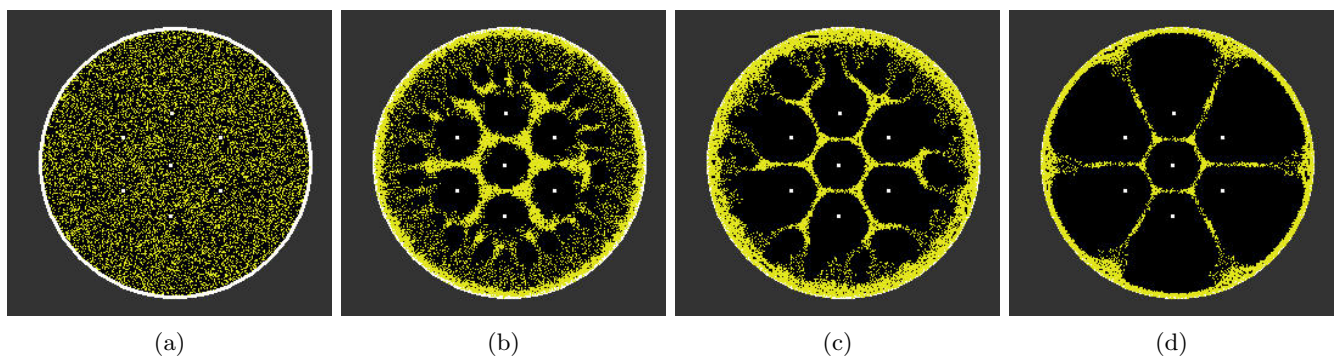

Fig. 10.   Approximation of Voronoi diagram by model in response to repulsive field. (a) Initial distribution of particles (yellow) representing a uniform mass of plasmodium, (b-c) particles respond to repulsive field by moving away from repellents, (d) final network connects outer attractant and bisectors correspond to Voronoi diagram.

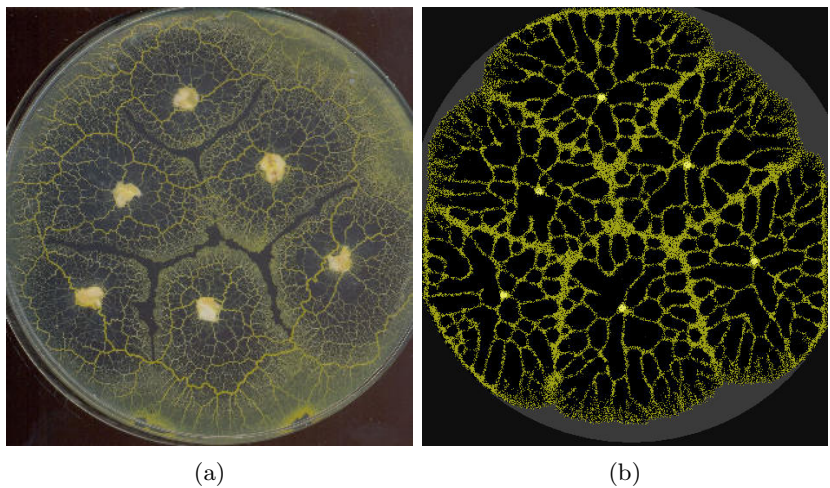

Fig. 11. Approximation of Voronoi diagram by merging method. (a) Approximation of Voronoi diagram by plasmodium of *Physarum polycephalum*. Plasmodia are inoculated on oat flakes onto oatmeal agar. Radial expansive growth of plasmodium is temporarily inhibited at regions where another plasmodia is occupied. This region indicates bisectors of Voronoi diagram, (b) Attempt to reproduce bisector formation by model. The model plasmodium (yellow, particle positions shown) is not inhibited at regions of fusion, but bisector position is indicated by the increase in network density at these regions.

The model plasmodium simultaneously grows outwards (drawn outwards by the surrounding attractant substrate) and is repelled from behind by the diffusion of the repellent field (Fig.12b,c). The Voronoi bisectors are initially coarsely approximated by the meeting growth fronts (Fig.12d) and the bisectors are then refined as the minimising network formed by the particle population avoids the repellent field from the planar shapes (Fig.12,e-f).

### 5.5. *Towards Hybrid Voronoi Diagrams*

The results in Figs. 10–12 demonstrate that the model can reproduce the approximation of the Voronoi diagram using both the repulsion method in [Shirakawa *et al.*, 2009] and the merging fronts method in [Adamatzky, 2010b]. The results also demonstrate approximation of the Voronoi diagram for more complex 2D shapes and the simultaneous use of attractant and repulsion fields for the computation.

Transport networks formed under the influence of only attractant stimuli take the form of minimising proximity graphs, specifically reproducing the range of graphs in the Toussaint hierarchy (see [Toussaint, 1980] and also [Adamatzky, 2008] for the same behaviour implemented by *Physarum*). At low nutrient concentration the graphs formed by the model minimise their initial configuration by self-organisation to approximate the Steiner tree and Minimum Spanning Tree, whereas at higher concentration the networks approximate Relative Neighbourhood Graphs and Gabriel graphs (as demonstrated in [Jones, 2010a]). How does the relative concentration of repellents affect the innate minimising behaviour of the networks?

An example of how differences in -ve concentration affect network structure is shown in Fig. 13. In this example the minimising particle network adapts to the presence and concentration of the repulsive field surrounding the planar shapes. In the first image the -ve stimuli overwhelms the contractile properties of the particle network and a planar Voronoi diagram is formed (Fig. 13a). As the concentration of -ve stimuli emanating from the shapes is reduced, however, the contractile behaviour of the particle network exerts its influence, shrinking the network around the shapes (Fig. 13b,c). At the lowest repellent concentration, the shapes are wrapped tightly by the 'band' of particles (Fig. 13d). Note that the internal network does not adopt the bisectors of the Voronoi Diagram, but instead adopts straight lines which form a Steiner point (S) at the location equidistant between the three sources (circled) connecting to the point.

The results in Fig. 13 illustrate the effects of a reduction in repellent concentration on Voronoi patterning as contractile forces begin to exert an effect. What is the response in the opposite situation where initially predominating contractile forces become outweighed by repellent concentration? Fig. 14 illustrates the situation where a set of point sources are individually surrounded by small contractile bands of parti-cles. At very low concentration, the pattern simply corresponds to the pattern of data sources (Fig. 14a). As the repellent concentration increases, however, the expansive force predominates and the cells expand (Fig. 14b) and merge with close neighbours forming clusters of neighbouring cells (Fig. 14c,d,e) until a single cluster of cells is formed.

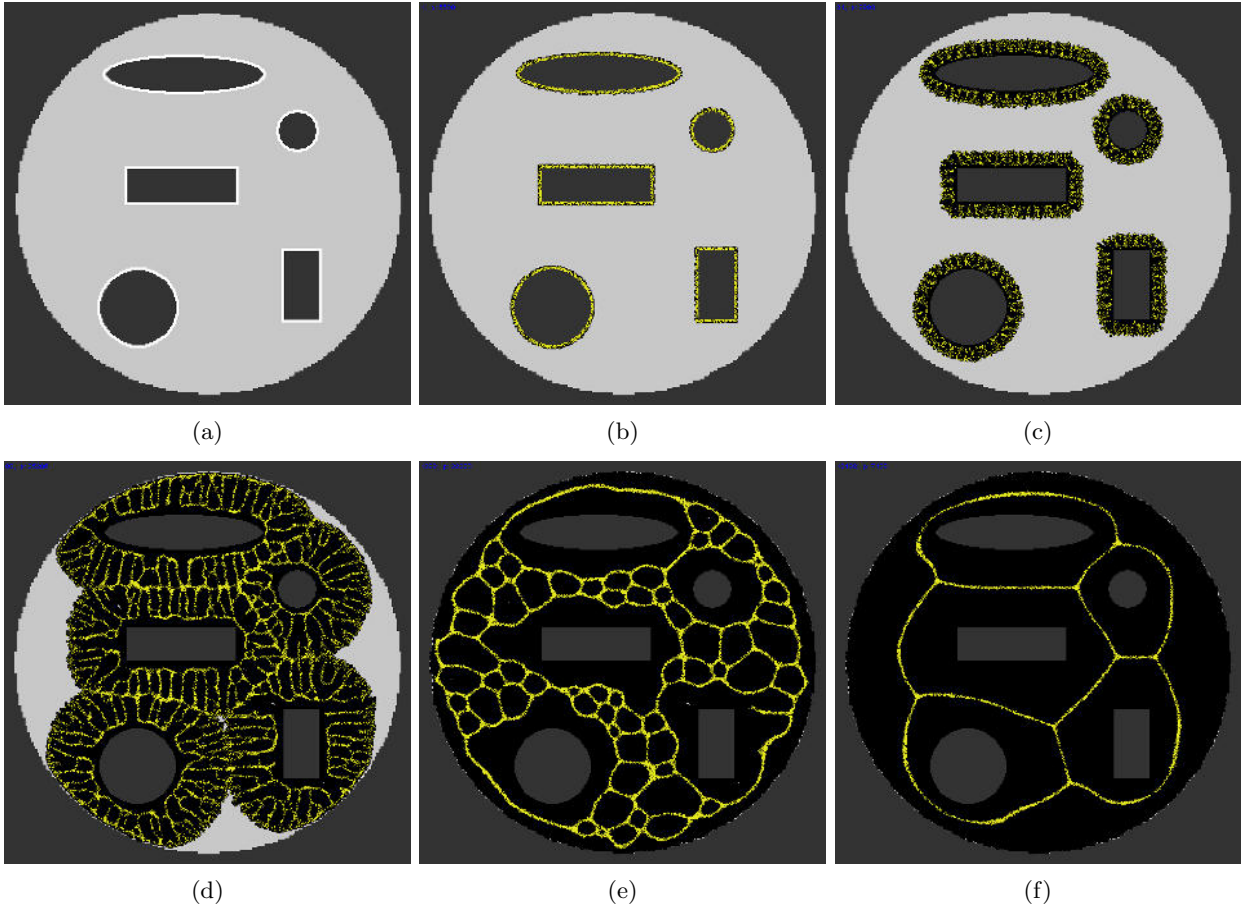

Fig. 12. Approximation of planar Voronoi diagram of complex shapes using combined repulsion and merging. (a) Configuration of arena showing inoculation borders of shapes (white), background substrate (light grey) and repulsive regions (dark grey), (b) model (yellow) is initialised at edges of shapes $t=1$, (c) model population extends outwards into arena attracted by gradient of nutrient substrate (light grey) $t=21$, (d) merging growth fronts represent coarse position of Voronoi Diagram $t=80$, (e) and (f) model network adapts to repulsive field from data sources and arena wall to form final approximation of Voronoi Diagram $t=1388$ and $t=12188$.

The cellular tessellation resembles, but does not perfectly match the Voronoi diagram. Fig. 15 illustrates differences between the cellular Voronoi diagram (Fig. 15a) and the 'full' Voronoi diagram (overlaid). The circled areas indicate regions where the bisector of the cellular diagram does not match the full diagram. The cause of this distortion can be seen in Fig. 15b which shows a map of the diffusion field including the contractile network trail (light shaded bisectors) and the gradients emanating from the point sources. It can be seen that in the areas corresponding to the circled regions in Fig. 15a, the contractile force of the network is stronger than the repellent diffusion sources and the bisectors are pulled from the Voronoi bisectors (overlaid). Why is the network pulled away from the outer points? This is partially due to the general nature of the contraction force (the tension forces ensure an inward bias) and partially due to the fact that smaller cells in minimising networks tend to shrink, whilst larger cells tend to grow. It is only at very high repellent concentration that the contractile force is overwhelmed by the repellent force and the Voronoi diagram is correctly approximated (Fig. 15c). Weighted Voronoi diagrams are generated in classical methods by modifying the distance measure during its computation (for example multiplying the distance by a weighting factor). This weighting may be approximated in the model by varying the relative size of the data points. This affects the amount of chemoattractant projected into the lattice and distorts the diagram to reflect the relative size of the data points within each cell ((Fig. 15c)).

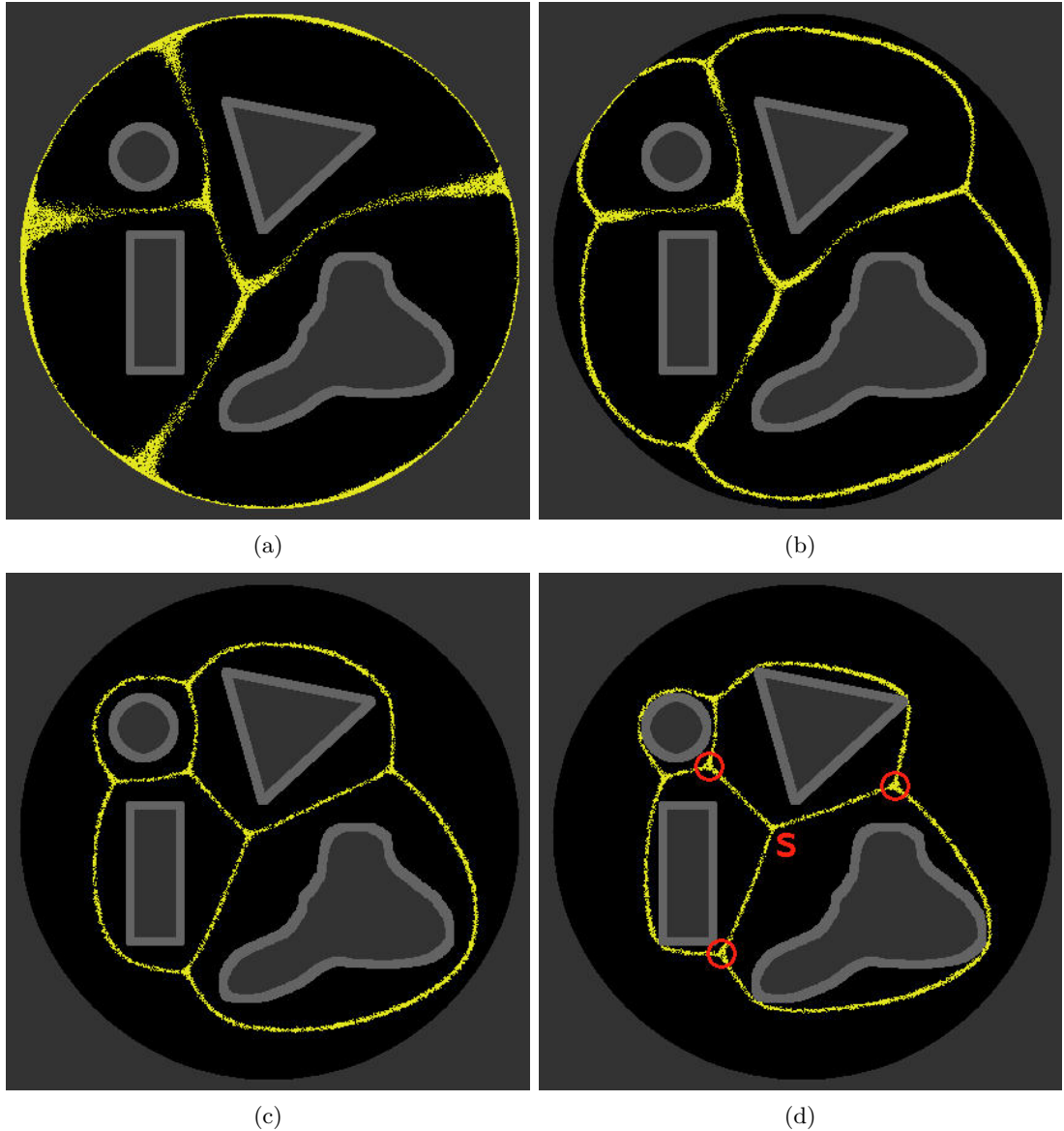

Fig. 13. Reducing repellent concentration allows minimising behaviour to exert its influence, inducing formation of hybrid Voronoi diagram. (a) at high concentration the repellent gradient forces the contractile network (yellow) to conform to the position of curved Voronoi bisectors between planar shapes, (b-d) reduction in repellent concentration allows contractile effects of transport network, minimising the connectivity between cells.

The combination of contractile particle networks with repulsive stimuli may be used as an unconventional approximation of geometry problems. In the paper by Kim et al. [Kim *et al.*, 2005] the problem of generating a Voronoi diagram of circles within a bounding circle was addressed using a multi-step process based on constructing a classical Voronoi diagram, generating a seed topology to represent the outer bounding circle, and generation of the exact edge locations. Can the exact solution described in [Kim *et al.*, 2005] and shown in Fig. 16a be approximated in an unconventional computation method? When initialised at random locations in the arena the contractile particle population self-assembled into a network surrounding the circles which represented the repellent sources (Fig. 16b). We found that the innate material behaviour was able to approximate closely (but not exactly) the multi-stage result obtained by Kim et al. Further reductions in repellent concentration maintained the separate partitioning of the circles whilst reducing the internal distance of the network (Fig. 16c). At very low repellent concentrations the circles exerted very little repellent force and merely became obstacles to the minimisation of the particle network, constraining its evolution. The final network encompasses all of the original nodes but appears to minimise the connectivity between the nodes (Fig. 16d).

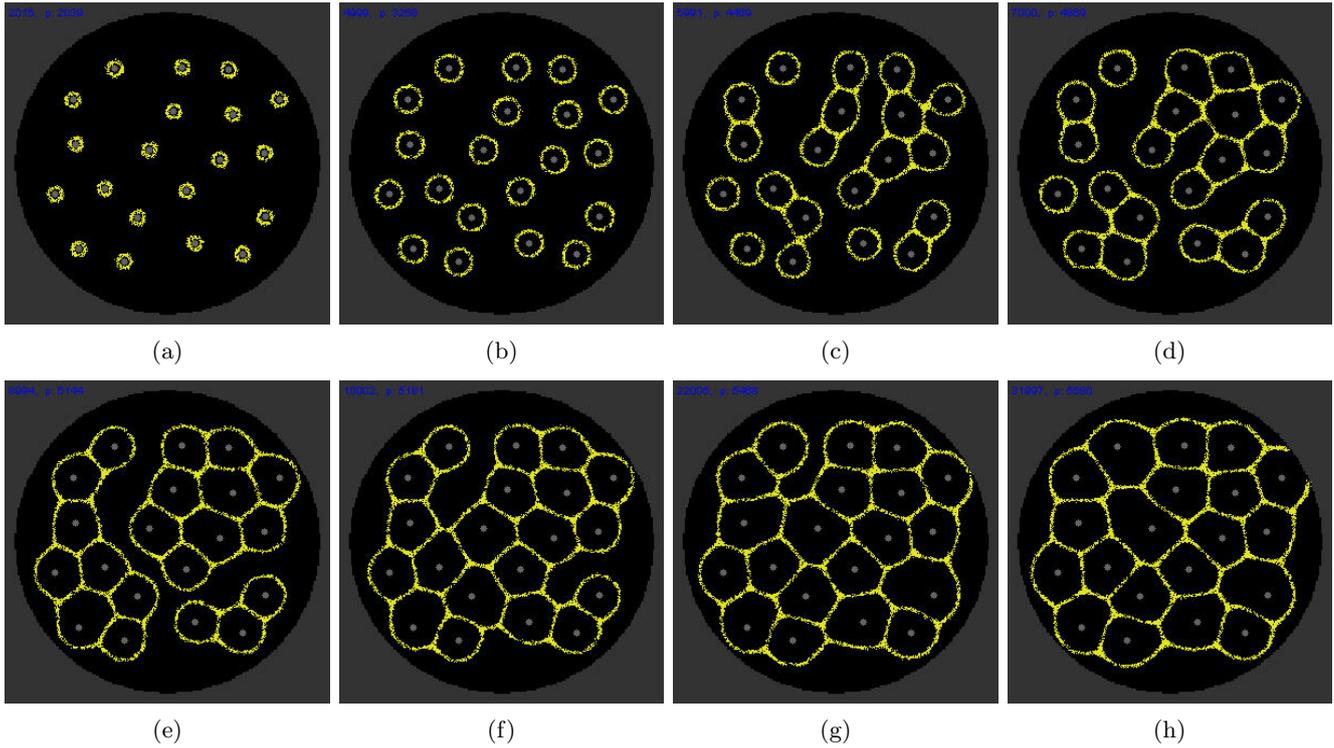

Fig. 14.  Increasing repellent concentration forms an hybrid 'cellular' Voronoi diagram. (a) contractile networks (yellow) initialised around 20 nodes with -ve stimuli emanating from nodes of weight 0.005, (b)-(h) increasing -ve stimuli strength expands networks and cell borders fuse, forming a hybrid 'cellular' Voronoi diagram. Node concentration 0.025, 0.05, 0.075, 0.125, 0.15, 0.45, 0.7 respectively.

When repellent stimuli are added to a plasmodial network already containing attractants (Fig. 17a), the plasmodium abandons its occupation of repellent regions as the repellent diffuses into the surrounding agar. The resulting network avoids the repellents whilst maintaining connectivity between the nutrient sources (Fig. 17b). When a model network is also presented with the addition of repellents, the connectivity between the attractant nodes is maintained and the network course adapts to avoid the repellent sources (Fig. 17c,d).

## 6.  Conclusions

We have examined the influences of repellent fields on the innate minimising behaviour of synthetic transport networks in the approximation of Voronoi diagrams in unconventional computation systems. Voronoi diagrams are the prototypical applications of unconventional computing architectures which utilise phenomena of parallel propagation of information to perform useful computation. In this paper we used a particle approximation of slime mould *Physarum polycephalum* which exhibits the same network formation and minimising adaptation as the organism itself and can be considered as a simple and distributed virtual computational material. We demonstrated how the synthetic particle networks can reproduce the experimentally observed approximation of Voronoi diagrams using *Physarum* using the repulsion method of [Shirakawa *et al.*, 2009], and the merging fronts method of [Adamatzky, 2010b].

By combining the innate minimising behaviour of the particle networks with repellent diffusion sources we were able to approximate a range of hybrid Voronoi diagrams. The network patterns were dependent on the shape and concentration of the repellent stimuli. High concentration stimuli reproduced Voronoi diagrams in both simple point sources, circular sources, growing cellular sources, and for complex planar shapes. Weighted Voronoi diagrams were approximated by varying the relative size of the data sources. Low concentration repellent stimuli reduced the propagation distance of the repulsive field and allowed the innate contraction behaviour of the particle model to exert an influence on network connectivity. The resulting hybrid networks encompassed the repellent sources, retaining their separate partitioning and minimising the internal network distance between the sources. The combination of separate partitioning of objects in the plane and the minimisation of the distance between the grouped objects may prove useful in a range of geometry problems, including packing, wiring and bundling problems, path planning and robotic guidance. When presented with a combination of attractant and repellent stimuli, the model reproduced the behaviour of the *Physarum* plasmodium by maintaining connectivity of the attractants whilst avoiding the repellent stimuli.

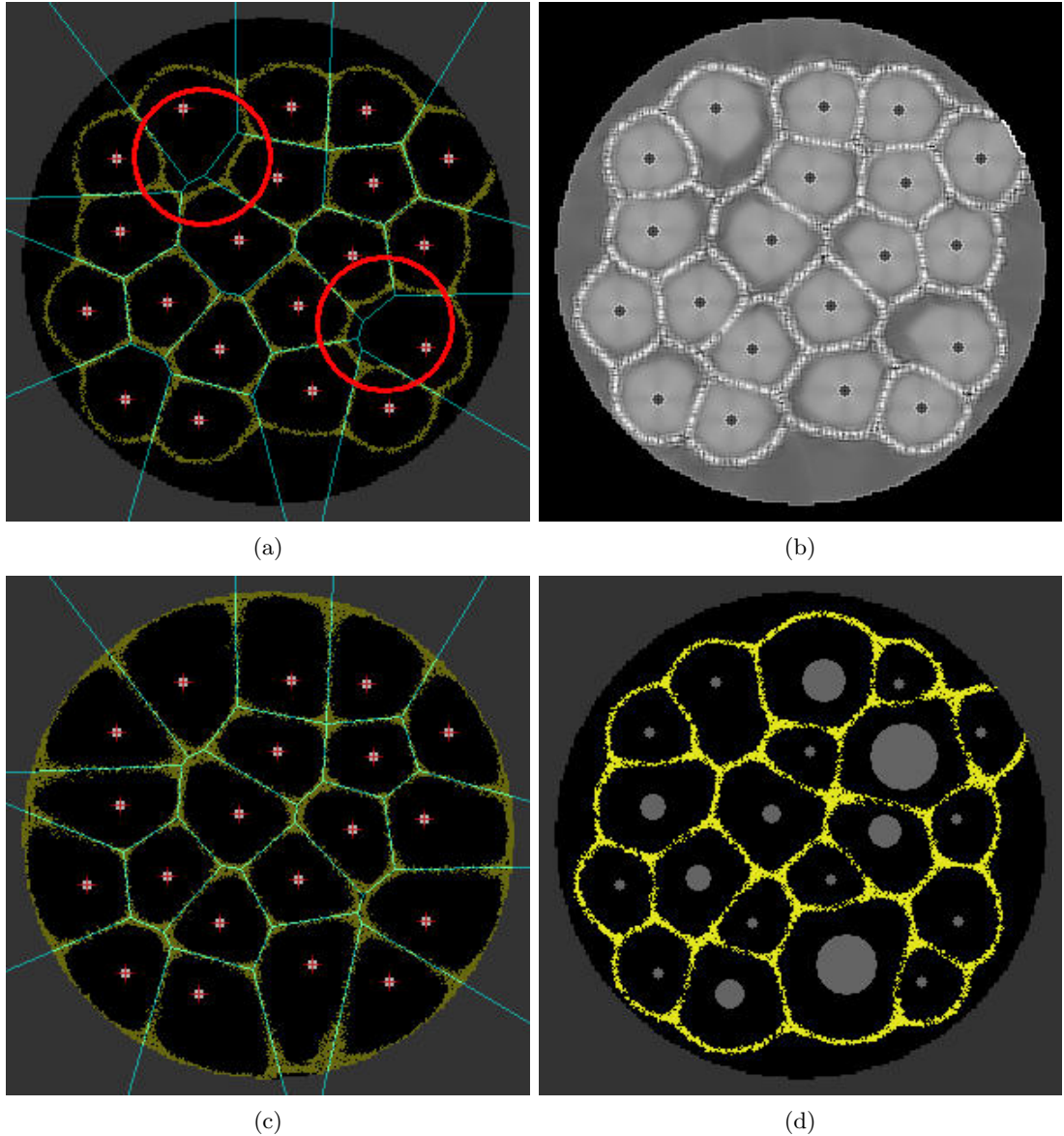

Fig. 15. Differences between 'cellular' and 'full' Voronoi diagram and approximation of weighted Voronoi diagram. (a) 'cellular' Voronoi diagram (yellow) formed by method in Fig. 14 showing differences (circled in red) between 'full' diagram (overlaid in blue), (b) visualisation of contractile network field and repellent field shows shaded regions where contractile network forces overcome repellent forces (see text), (c) particle network (yellow) perfectly matches full Voronoi diagram (blue) only at very high -ve stimuli node concentration (Voronoi diagram by classical method is overlaid), (d) Weighted diagram is approximated by varying size of source data points.

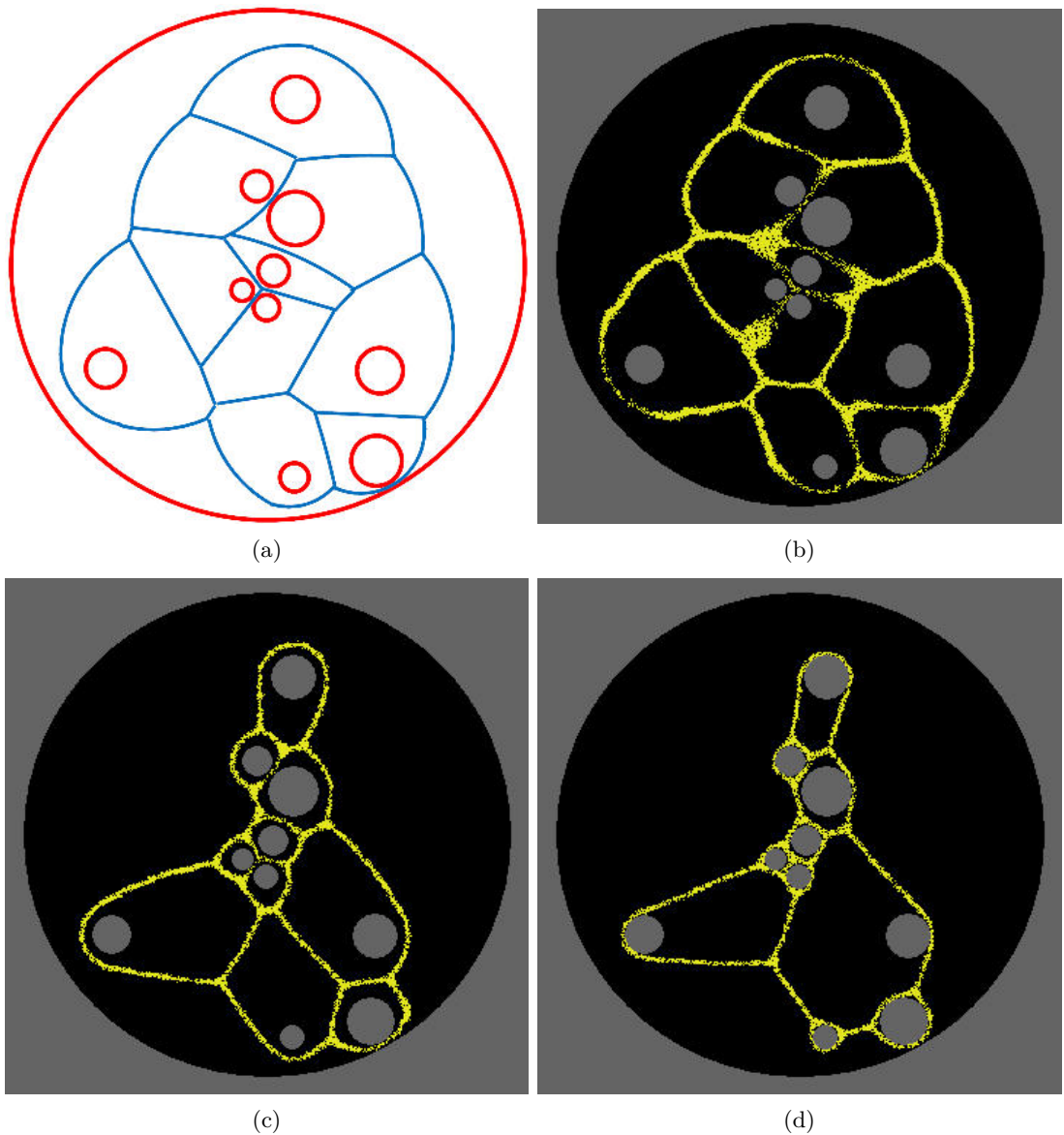

Fig. 16.  Approximation of the circle Voronoi Diagram problem and its subsequent minimisation using unconventional computation approach. (a) Circle Voronoi diagram as proposed in [Kim *et al.*, 2005], generating circles shown in red and Voronoi diagram in blue (b) approximation of the circle Voronoi diagram in model plasmodium (yellow) at high repellent concentration, (c-d) reducing repellent concentration maintains partition of shapes whilst reducing network connectivity.

Can the response of synthetic particle networks to +ve and -ve stimuli be rationalised in terms of the behaviour of the organism itself? For attractant stimuli the benefits of minimal network connectivity are obvious and well documented [Nakagaki *et al.*, 2004], [Adamatzky, 2008]. For repellent stimuli the avoidance of hazards in the environment are also advantageous [Shirakawa *et al.*, 2009], [Shirakawa & Gunji, 2010], [Adamatzky, 2009b], [Adamatzky, 2010d]. By combining the minimal connectivity to (for example) nutrients whilst avoiding hazards in the environment the plasmodium can increase its chances of obtaining energy whilst minimising risk of damage. It is notable that the organism's trade-off between reward and risk in its environment is 'computed' by the response of the very material of which it is composed. This has the advantage of not requiring any neural integration of separate +ve and -ve stimuli. The complexity of slime mould's response, and the rich variety of partitioning and network connectivity in the synthetic particle model to similar stimuli, may suggest ways in which artificial computing schemes can take advantage of opposing signalling cues by means of innate material responses.

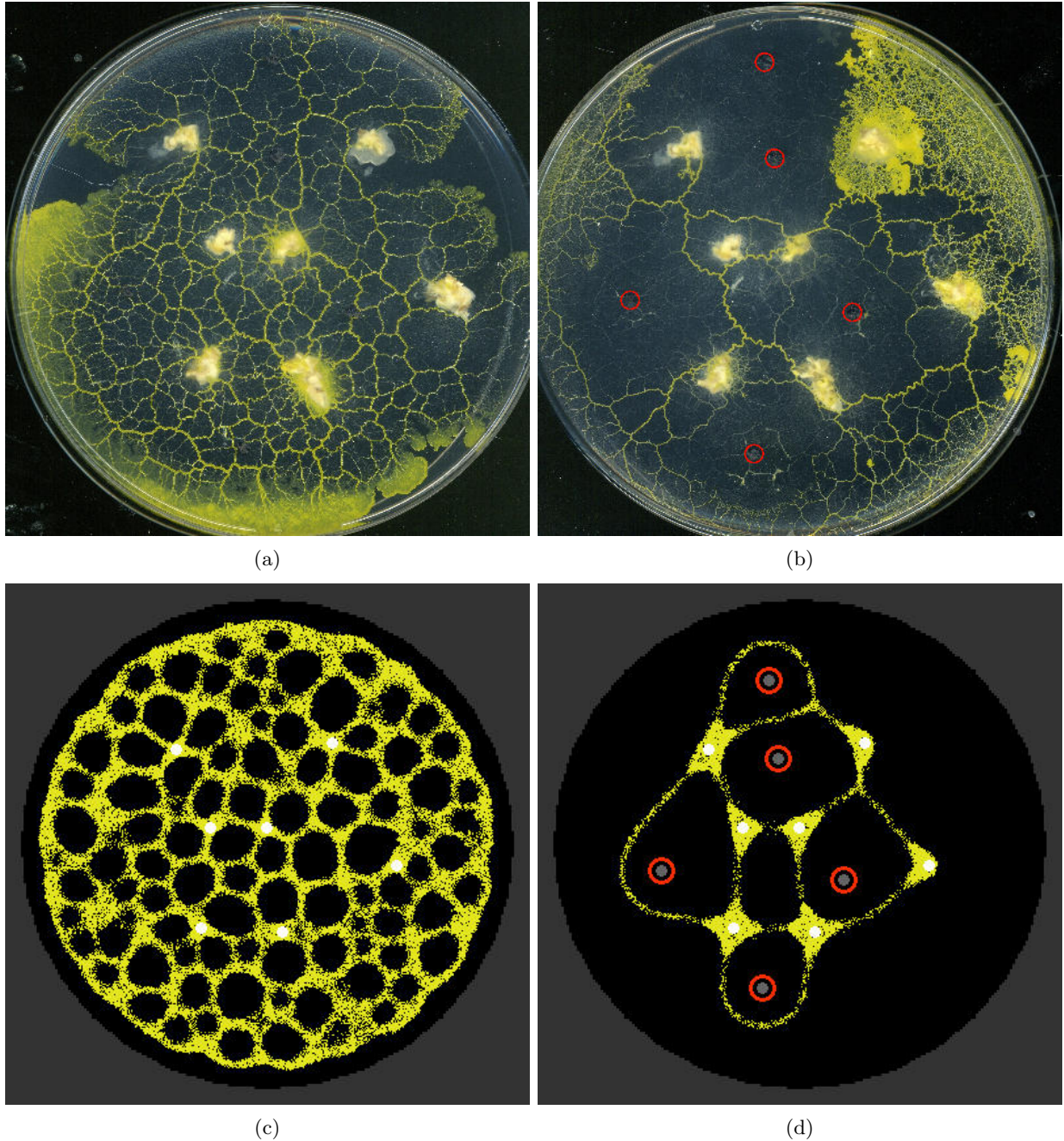

Fig. 17. Addition of repellents to attractants forms hybrid of Voronoi diagram and proximity graph. (a) protoplasmic network formed in the presence of nutrient oat flakes, (b) addition of repellent (salt crystals, circled in red) causes plasmodium to abandon repellent areas whilst maintaining connectivity with nutrients after 12h, (c) initial simulated protoplasmic network in model (yellow), (d) response of model to the presence of repellents connects the attractants (white) and avoids repellents (circled in red).

## Acknowledgments

This work was supported by the EU research project "Physarum Chip: Growing Computers from Slime Mould" (FP7 ICT Ref 316366)